\documentclass[sigconf,nonacm,natbib=false]{acmart}

\AtBeginDocument{%
  }

\RequirePackage[
  datamodel=acmdatamodel,
  style=acmnumeric,
  ]{biblatex}

\usepackage{soul}
\usepackage{subcaption}
\usepackage{stfloats} 

\begin{document}

\title{Fake It No More: Evaluating L4S with SCReAM on Video Traffic}

\author{Nawel Alioua}
\affiliation{%
  \institution{UC Santa Barbara}
  \country{}
}
\email{nawel@ucsb.edu}

\author{Ryan Zanone}
\affiliation{%
  \institution{UC Santa Barbara}
  \country{}
}
\email{ryanzanone@ucsb.edu}

\author{Cheng Xi}
\affiliation{%
  \institution{UC Santa Barbara}
  \country{}
}
\email{cxi@ucsb.edu}

\author{Elizabeth Belding}
\affiliation{%
  \institution{UC Santa Barbara}
  \country{}
}
\email{ebelding@ucsb.edu}

\begin{abstract}
The growing interest in Low Latency, Low Loss, and Scalable Throu\-ghput (L4S) reflects the need for lower latency in interactive multimedia applications. In this paper, we use an open-source DualPI2 implementation over the Mahimahi emulator to evaluate the impact of L4S on SCReAM congestion controlled video traffic. To do so, we augment the SCReAM BW tool with a video codec, enabling the generation of video traffic in addition to its original synthetic RTP mode. We evaluate both network-level and Quality of Experience (QoE) metrics on a mobile network trace, under random packet loss, and with different motion-complexity levels. In our baseline scenario, L4S reduces the median per-run $95^{th}$ percentile queue delay by 35\%, at the cost of a 42\% drop in sender throughput. Under 1\% packet loss, L4S yields more pronounced QoE gains compared to the lossless scenario, despite narrower network-level benefits. Across video content complexities, L4S also maintains more stable QoE than Classic. These results underscore the importance of evaluating QoE alongside network-level metrics when assessing the effect of L4S on end-user application performance.
\end{abstract}



\maketitle

\section{Introduction}
The increasing demand for low-latency and reliable real-time communication services has driven the development and adoption of the Low Latency, Low Loss, and Scalable Throughput (L4S) architecture~\cite{comcast-l4s-2025, comcast-xfinity}. L4S, described in RFC 9330~\cite{rfc9330}, aims to provide low latency while maintaining high throughput. L4S uses Explicit Congestion Notification (ECN) to signal congestion early, thereby avoiding packet drops and enabling finer-grained rate adaptation through scalable congestion control at the end hosts. To facilitate this behavior, the bottleneck isolates L4S and Classic traffic into separate queues and signals queue growth promptly via ECN~\cite{rfc9331, rfc9332}.

At a high level, the L4S architecture relies on two main components: a L4S-compatible Active Queue Management (AQM) at the network bottleneck, and scalable congestion control at the end hosts. Dual-queue coupled AQMs, as described in RFC 9332~\cite{rfc9332}, differentiate between Classic and L4S traffic, providing low latency for the latter while ensuring fair coexistence with the former. The DualPI2 AQM~\cite{olga-dualpi2} is the de facto implementation of such a dual-queue coupled AQM in the Linux kernel. More recently, an implementation of DualPI2 within the Mahimahi emulator~\cite{mahimahi-atc, dualpi2-mahimahi-paper} became available~\cite{mahimahi-dualpi2-github-repo}, making controlled L4S experimentation more accessible to the research community.

As a scalable congestion control protocol, Self-Clocked Rate Adaptation for Multimedia (SCReAM)~\cite{scream-paper, rfc8298} is designed for real-time multimedia traffic. It operates based on the self-clocking principle, aiming to send data at a rate comparable to the exit rate from the network, and is specifically optimized for mobile networks where variability and channel jitter are high~\cite{scream-paper}. SCReAM utilizes the Real-time Transport Protocol (RTP) to carry multimedia traffic, and expects feedback over the Real-time Transport Control Protocol (RTCP) to adjust its congestion window and sending rate. It supports non-ECN traffic; "classic" ECN where the congestion signal used is the ECN Congestion Experienced (CE) mark rather than the traditional packet drop; and L4S, where the rate reduction is proportional to the fraction of ECN CE-marked packets. SCReAM has been standardized as RFC 8298~\cite{rfc8298} in 2017, and work is ongoing on its updated version SCReAMv2~\cite{johansson-ccwg-rfc8298bis-screamv2-07}.  

SCReAM is available as an open-source project~\cite{scream-github-repo}, integrated within a bandwidth probing tool (the BW tool). However, this tool only supports fake RTP traffic generation. While this capability is valuable for early-stage experimentation and for debugging under various network conditions, it poses a limitation for more realistic traffic testing, such as Quality of Experience (QoE) evaluation of video traffic. More broadly, to assess not only SCReAM itself but also the L4S paradigm end-to-end, experiments should capture the interaction between congestion control and realistic multimedia traffic generation. This work extends the BW tool with video encoding support to enable such experiments with actual video traffic. We then use this extension to evaluate SCReAM over Classic and L4S modes in a Mahimahi-based setup, considering both network-level and QoE metrics. Enabling such traffic generation in the BW tool therefore provides the community with a practical platform for studying L4S performance on multimedia applications.

\vspace*{0.1in}
Our main contributions are the following:

\begin{itemize}
  \item We extend the SCReAM BW tool with encoded video traffic capabilities, moving beyond its original synthetic-traffic operation and enabling controlled experiments with actual video content.
  
  \item We compare SCReAM in Classic and L4S modes on Mahima\-hi emulated DualPI2 AQM using both synthetic and actual video traffic, highlighting how application-level video \allowbreak behavior complements network-level observations  from synthetic traffic.
  
  \item We show that L4S improves key network and playback-related metrics, compared to Classic traffic mode, and remains relatively stable across video content of different motion complexity, while also examining robustness under random loss.

  \item We publicly release the encoder-augmented BW tool to support reproducible evaluation of SCReAM congestion-controlled video traffic\footnote{https://github.com/5G-VCA-CC/scream}.

\end{itemize}

\section{Integrating Video Encoding into the SCReAM BW Tool}
In this section, we describe the design and implementation choices made while integrating a video codec into the SCReAM BW tool. Our goal is to move beyond the tool's original synthetic RTP traffic generation and enable controlled experiments with encoded video traffic, while preserving the BW tool's role as a lightweight and reproducible evaluation platform. For this task, We took inspiration from the Ringmaster project~\cite{ringmaster-github-repo}, which not only includes encoder and decoder components, but also incorporates several practical features that improve streaming robustness and make it a useful reference point as a lightweight tool for video streaming and conferencing research~\cite{tambur}.

\subsection{Video codec}
We use VP9~\cite{vp9-paper}, an open-source video codec introduced by Google and implemented in \texttt{libvpx}~\cite{libvpx-code}. VP9 is a suitable choice for our integration because it is widely available, open source, and supported in WebRTC implementations~\cite{mdn_webrtc_codecs}. The general design of the encoder and decoder modules was modeled after Ringmaster, while introducing appropriate adaptations to the SCReAM- and BW tool-specific hook points and module interfaces. 

\subsection{Main interactions with the SCReAM congestion controller}
The integration of VP9 into the SCReAM BW tool mostly follows a pull-based interaction model, where the sender and encoder query the SCReAM API and then act according to the returned values. However, the codec also passes information back to SCReAM to maintain accurate controller state. We summarize the main interactions below.

\vspace*{0.1in}
\noindent \textbf{SCReAM to codec signals.} In our implementation, these signals are exposed as return values from SCReAM API calls rather than as callbacks.

\vspace*{0.05in}
\noindent \textit{Target bitrate and packetization guidance:} The main interface between the encoder and SCReAM is the transmission of the estimated target bitrate. SCReAM provides the target bitrate through the \texttt{getTargetBitrate()} function and the Maximum Segment Size (MSS) through the \texttt{getRecommendedMss()} function. These two values are then passed to the encoder to guide the encoding and packetization of the next frame.

\vspace*{0.05in}
\noindent \textit{Loss indication:} SCReAM can also inform the encoder about packet loss occurring on the medium through the \texttt{isLossEpo\-ch()} function. We use this signal, in addition to the case where \texttt{getTargetBitra\-te()} returns a negative value, to request a keyframe from the encoder. We discuss keyframe injection mechanisms in more detail in \S\ref{sec:keyframe}.

\vspace*{0.1in}
\noindent \textbf{Codec to SCReAM signals.} The reverse direction of the interaction consists of state updates from the media pipeline back to SCReAM. For every RTP packet produced as part of an encoded frame, the encoder calls the \texttt{newMediaFrame()} function as a media-production notification. It also uses the \texttt{addTransmitted()} function after every actual RTP packet transmission to update SCReAM's state, ensuring that the congestion controller tracks the packets that have effectively left the interface.

On the decoder end, the SCReAM function \texttt{receive()} is called upon reception of an RTP packet, thereby updating SCReAM's view of packet arrivals at the receiver. The function \texttt{createStandardi\-zedFeedback()} is called when an RTCP feedback message is about to be formed, allowing the receiver to generate the feedback required to close the congestion-control loop.

\subsection{Keyframe injection}
\label{sec:keyframe}
We introduce several keyframe generation techniques that can be enabled using command-line options, some of which can be combined. To avoid excessive keyframe generation, we do not request more than one keyframe within any 100 ms interval, following a similar implementation in the SCReAM source code.\footnote{\url{https://github.com/EricssonResearch/scream/blob/master/code/wrapper_lib/screamtx_plugin_wrapper.cpp\#L859}} With the exception of the periodic mode, these mechanisms can be enabled simultaneously.

\vspace*{0.1in}
\noindent \textbf{Keyframes triggered by target bitrate estimation.} If the \texttt{getTar\-getBitrate()} function returns a negative value, a keyframe is requested immediately from the encoder as recommended in the codebase.\footnote{\url{https://github.com/EricssonResearch/scream/blob/master/code/ScreamTx.h\#L329}} 

\noindent This function returns \texttt{-1} when SCReAM detects a severe recovery condition, such as an RTP queue discard event, an ACK received outside the window and therefore treated as lost forever, or a late ACK, also treated as lost. Importantly, the function emits \texttt{-1} as a one-shot recovery signal rather than as a persistent rate value. After that first emission, subsequent calls return the regular target rate again until a new loss/discard episode occurs.

\vspace*{0.1in}
\noindent \textbf{Keyframes triggered by loss.}
If the \texttt{isLossEpoch()} function returns \texttt{true}, then we wait for 100 ms before forcing a keyframe. \texttt{isLossEpoch()} returns \texttt{true} when packet loss causes a nonzero lost-rate update, or when SCReAM discards packets from an overgrown RTP queue. After a loss is detected, the function clears the internal flag so repeated calls return \texttt{false} until another new loss epoch is detected. 
\newpage

\noindent \textbf{Keyframes triggered by long-standing unacked packets.} This mechanism was ported from Ringmaster~\cite{ringmaster-github-repo}. In this case, a keyfra\-me is triggered if the oldest transmitted packet has remained unacknowledged for at least one second. This process also triggers the clearing of the transmitted packet history up to that point.

\vspace*{0.1in}
\noindent \textbf{Periodic keyframes.} This mode generates periodic keyfra\-mes every $n$ seconds, provided as a command-line parameter. Contrary to the existing periodic keyframe generation in the original BW tool, we do not prescribe a fixed multiplier for the size of the keyframe. Instead, we let the encoder decide what size a keyframe can reach based on the instantaneous target bitrate signal obtained from the congestion controller.

\subsection{Extra features}
\noindent \textbf{Packet retransmission.}
A smooth video playback requires minimizing packet loss, and retransmission is a key component in achieving this. To support retransmissions, we change the underlying implementation of the RTP queue data structure to a deque, which allows packets that need to be retransmitted to be queued at the front, effectively prioritizing them over newly generated packets.

\vspace*{0.1in}
\noindent \textbf{RTP header extension.}
We introduce additional RTP header fields for reconstruction of the video flow, consisting of a keyframe indication, the fragment count, and frame and fragment identifiers. These RTP extension fields let the receiver identify frame boundaries and fragment ordering explicitly, so the decoder can safely skip incomplete or lost frames up to the next complete keyframe.

\begin{figure*}[b] 
\centering
    \begin{subfigure}[b]{0.48\textwidth}
        \centering
        \includegraphics[width=\textwidth]{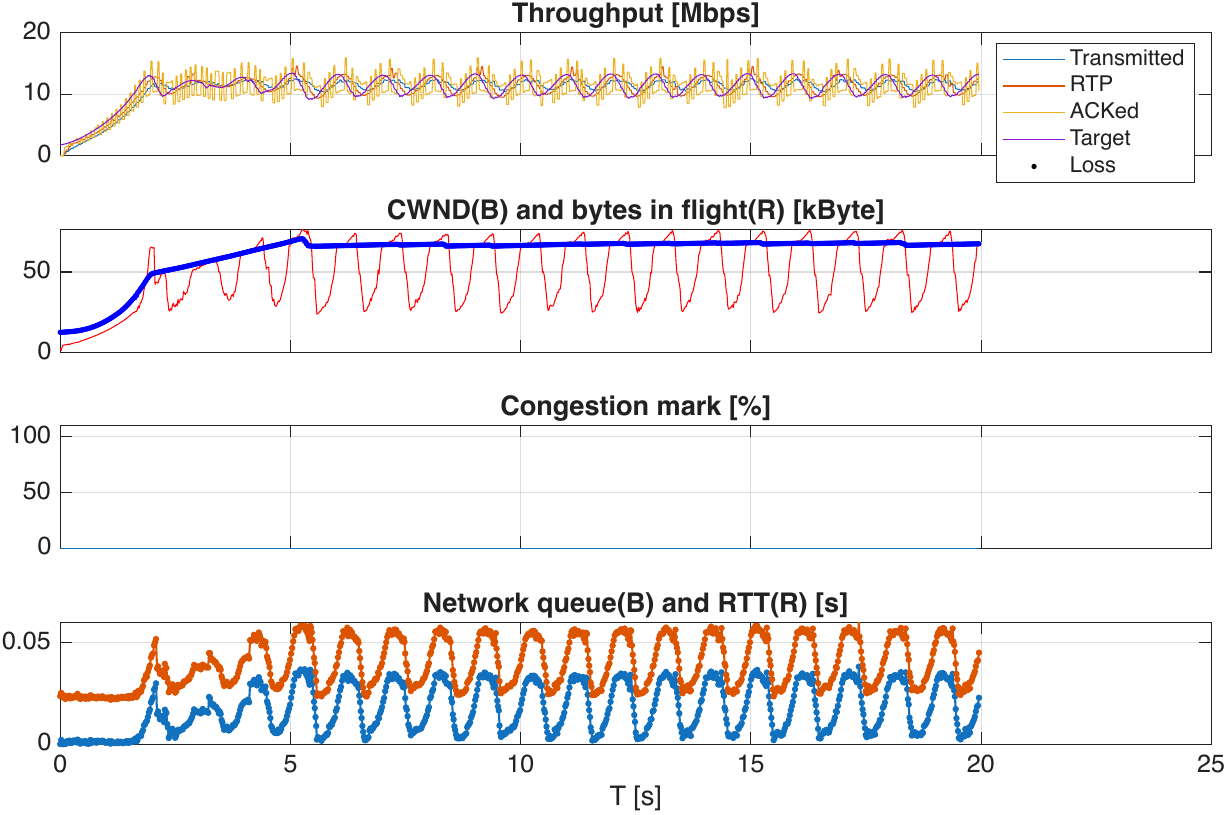}
        \caption{classic}
    \end{subfigure}
    \hfill
    \begin{subfigure}[b]{0.48\textwidth}
        \centering
        \includegraphics[width=\textwidth]{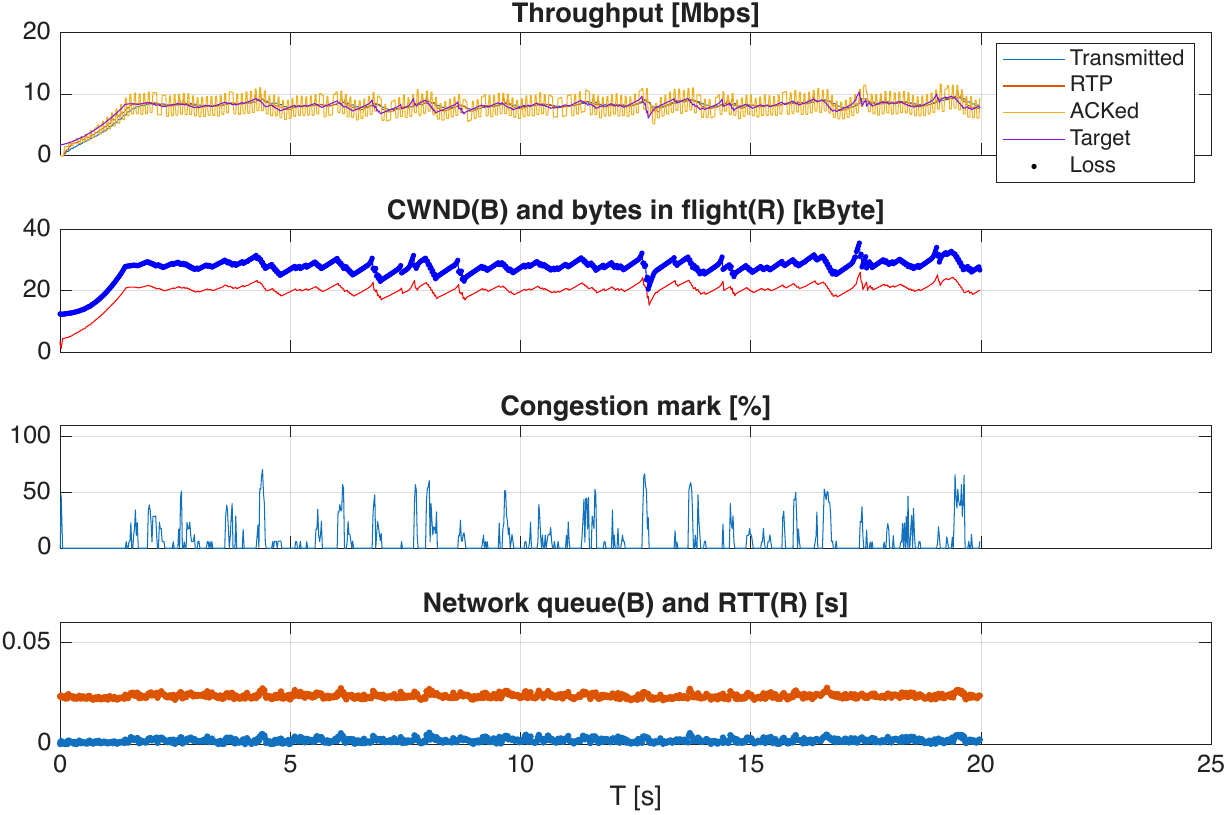}
        \caption{L4S}
    \end{subfigure}
    
\caption{Classic vs. L4S time-series metrics running synthetic RTP flows over Mahimahi DualPI2}
\label{fig:sanity-check}
\end{figure*}

\section{Experiment Setup \& Methodology}
This section describes the experimental setup and methodology used to evaluate how L4S affects SCReAM video traffic relative to Classic mode. We begin by presenting the Mahimahi-based emulation environment and the DualPI2 parameters used in our experiments. We then describe the evaluation metrics and video samples used to characterize application performance, followed by additional execution details, including the runtime environment and run configuration.

\subsection{Emulation system}
\label{sec:emulator}
We use the Mahimahi emulator~\cite{mahimahi-atc}, a widely used tool in networking and multimedia systems research for its high configurability and ability to conduct experiments under diverse network scenarios. We use the Mahimahi DualPI2 AQM implementation~\cite{dualpi2-mahimahi-paper, mahimahi-dualpi2-github-repo} to emulate the dual-queue bottleneck. Mahimahi provides a number of network traces by default; for our experiments, we select the short Verizon LTE trace. Table~\ref{tab:trace_stats} summarizes the characteristics of this uplink trace, obtained by running \texttt{iperf3} for the entire trace duration (140 seconds) using UDP to saturate the link, and logging the receiver throughput at 1 s intervals. The trace presents a considerable range of throughput values, with bouts of completely interrupted connectivity (0.0 Mbps). These conditions allow us to stress-test the SCReAM congestion-controlled video flow in a fluctuating network environment.

\begin{table}[htbp]
 \vspace{-5pt}
  \caption{Per-Second throughput statistics of the Mahimahi uplink Verizon-LTE-short trace (Mbps)}
  \label{tab:trace_stats}
  \centering
  
  \begin{tabular}{lccccc}
    \toprule
    Trace & Duration (s) & Min & Max & Avg. & SD \\
    \midrule
    Verizon  & 140 & 0.00 & 12.90 &  5.81 & 3.07 \\
    \bottomrule
  \end{tabular}
 \vspace{-5pt}
\end{table}

To validate the behavior of the Mahimahi DualPI2 AQM in the presence of L4S and Classic traffic, we perform a sanity check by running the SCReAM BW tool in its original fake RTP traffic mode through a Mahimahi setting with a fixed-rate trace of 12 Mbps and a Round-Trip Time (RTT) of 20 ms, for each of the L4S (\texttt{-ect 1}) and Classic modes (no \texttt{-ect}). The parameters of the DualPI2 AQM used in this experiment, and throughout the rest of this work, are detailed in \S\ref{sec:dualpi2-param}. The results depicted in Fig.~\ref{fig:sanity-check} show time-series plots of key metrics over a single 20-second run, including throughput, queue delay, and RTT for both modes. The behavioral differences we observe between L4S and Classic are similar in type and scale to those reported in the main BW tool repository~\cite{scream-github-repo}, supporting the use of the Mahimahi DualPI2 module for further experimentation.

\subsection{DualPI2 parameters} 
\label{sec:dualpi2-param}
We use the same default DualPI2 parameters as stated in RFC 9332~\cite{rfc9332}, namely $target=$ 15 ms,  $step\_thresh=$ 1 ms, $tupdate=$ 16 ms, $alpha=$ 0.16, and $beta=$ 3.20. In addition, we set $limit= $ 100 packets.

\subsection{Evaluation metrics}
\label{sec:metrics}
We collect execution logs consisting of a 2-second periodic output at the sender and the receiver, as well as a final per-run summary. Then, we extract the following metrics:

\vspace*{0.05in} 
\noindent \textbf{Transmission and reception (Tx/Rx) rates:} Actual sender transmit bitrate and the measured receive throughput.

\vspace*{0.05in} 
\noindent \textbf{Queue delay:} Estimated one-way queuing delay.

\vspace*{0.05in} 
\noindent \textbf{Packet loss:} Sender observed loss from feedback/loss detection as a percentage of total throughput.

\vspace*{0.05in} 
\noindent \textbf{Inter-frame (IF) delay:} Time between two consecutive completed frames at the receiver.

\vspace*{0.05in} 
\noindent \textbf{Freeze count:} Following the definition in WebRTC statistics~\cite{w3c-stats}, a \emph{freeze} is counted when: \[
\Delta t > \max\!\left(3 \times \bar{T}_f,\; \bar{T}_f + 150\right),
\]
where \(\Delta t\) is the gap in ms between two consecutive rendered frames, and \(\bar{T}_f\) is the average frame duration in ms.

\vspace*{0.05in} 
\noindent \textbf{Freeze duration:} Cumulative freeze duration of rendered frames experiencing a freeze as per the above definition.

\subsection{Video samples}
Video content differs in the level of motion complexity, generally quantified as Temporal Information (TI), which is defined as the amount of temporal changes in a video sequence between one frame and the next~\cite{itu-doc}. To study the effect of motion complexity on the performance of L4S and Classic video flows, we consider low-, medium-, and high-motion levels, and pick one 10-second video sample for each category.\footnote{Xiph.org Test Media Collection. \url{https://media.xiph.org/video/derf/}}

To verify the motion complexity level of each representative video, we use the \texttt{ffmpeg} tool and log a per-frame TI value for each considered video. The statistics summarized in Table~\ref{tab:video_stats} confirm that the selected clips span increasing levels of temporal complexity, from low to high motion.

\begin{table}[htbp]
  \caption{Per-frame Temporal Information (TI) for three sample videos}
  \label{tab:video_stats}
  \centering
  \small
   \vspace*{-0.01in}
  \begin{tabular}{lccccc}
    \toprule
    Sample Video & Level & Min & Max & Avg. & SD \\
    \midrule
    \texttt{akiyo\_cif.y4m} & Low & 0.95 & 6.80 & 2.99 & 1.37 \\
    \texttt{news\_cif.y4m}  & Medium & 1.10 &  35.21 & 7.86 & 4.6 \\
    \texttt{foreman\_cif.y4m} & High & 4.04 & 42.22 & 17.20 & 8.55 \\
    \bottomrule
  \end{tabular}
   \vspace*{-0.1in}
\end{table}

\subsection{Experimental Procedure}

Our experiments are conducted on a 14-core Intel Core Ultra 5 225H system with 64 GB of RAM running Ubuntu 24.04. Because the Mahimahi trace described in \S\ref{sec:emulator} lasts 140 seconds, we loop each video sample as needed to span the full trace and capture all network fluctuations. For each experiment, we collect data from 50 such 140-second runs. In the video-based experiments, we combine three keyframe generation techniques: target bitrate-based, loss-based, and unacked-packet-triggered, while enforcing the condition stated in \S\ref{sec:keyframe} that no two keyframes are generated within any 100 ms period.

\section{Experiment Results}
This section presents the results of three different experiments in terms of the metrics described in~\S\ref{sec:metrics}. Our goal in the first experiment is to understand and quantify the effect of L4S on the performance of a video flow in a mobile network scenario with an average RTT of 20 ms. In the second experiment, we examine the effect of introducing a random emulated packet loss on said performance. In the last experiment, we investigate how the video motion complexity  influences the observed performance differences. 

\subsection{Main test case: Medium motion complexity video}
\begin{figure}[ht]
\centering
    \begin{subfigure}[b]{0.323\columnwidth}
        \centering
        \includegraphics[width=\linewidth]{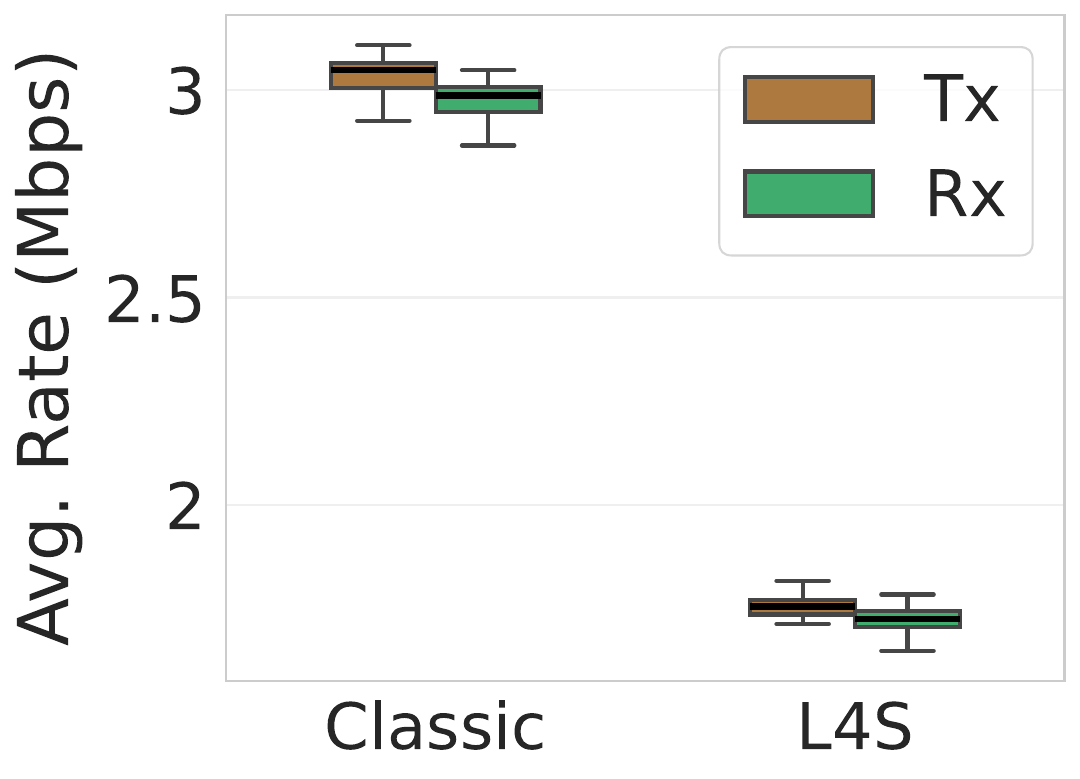}
        \captionsetup{font=footnotesize}
        \caption{Tx/Rx Rate}
        \label{txrx-medmo_verizon_loss-0_rtt-20ms}
    \end{subfigure}
    \begin{subfigure}[b]{0.323\columnwidth}
        \centering
        \includegraphics[width=\linewidth]{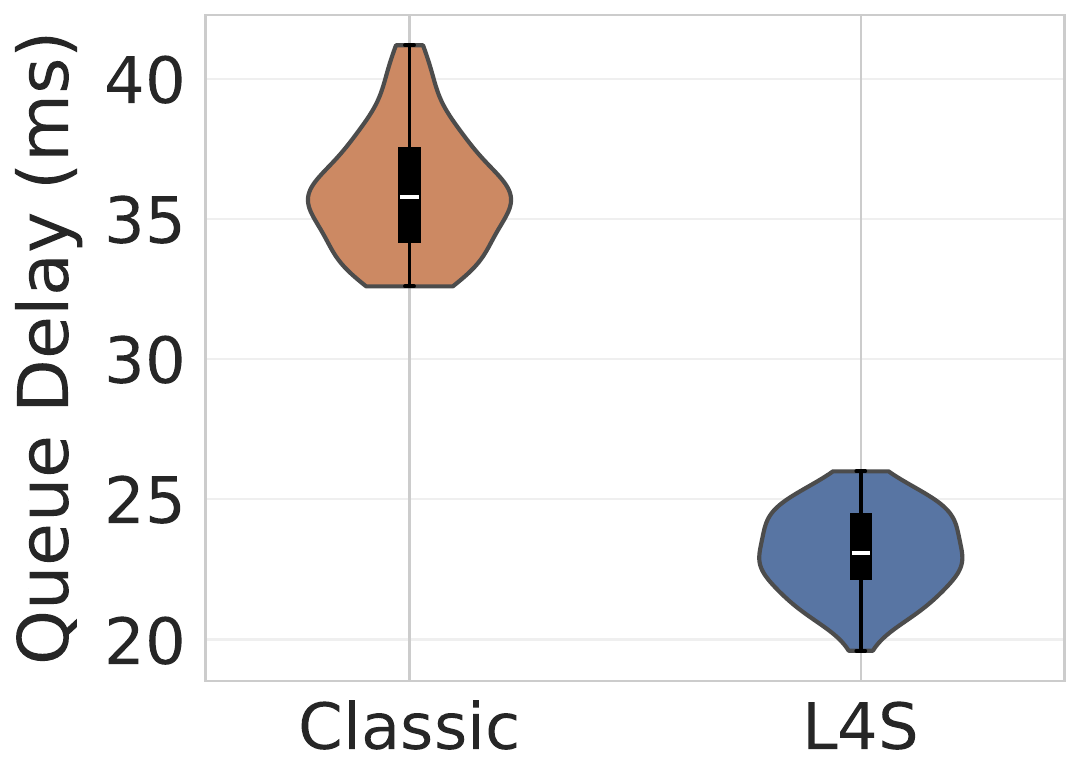}
        \captionsetup{font=footnotesize}
        \caption{Queue Delay (P95)}
        \label{qd-medmo_verizon_loss-0_rtt-20ms}
    \end{subfigure}
    \begin{subfigure}[b]{0.323\columnwidth}
        \centering
        \includegraphics[width=\linewidth]{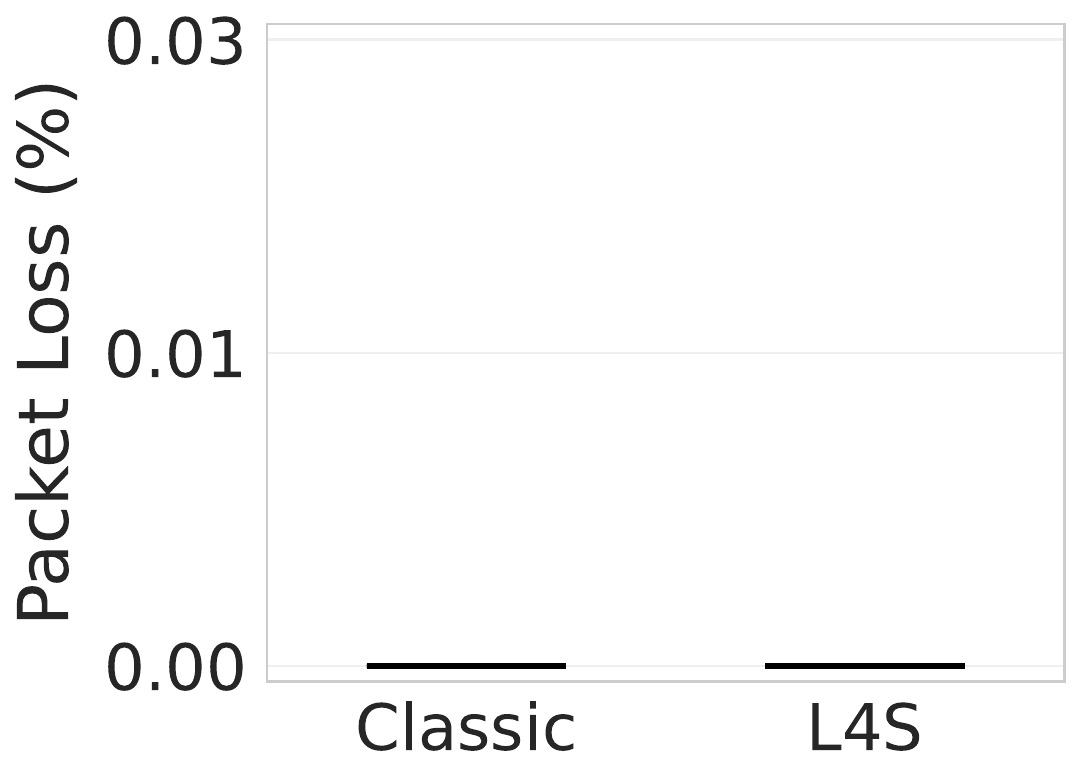}
        \captionsetup{font=footnotesize}
        \caption{Packet Loss}
        \label{loss-medmo_verizon_loss-0_rtt-20ms}
    \end{subfigure}


    \begin{subfigure}[b]{0.323\columnwidth}
        \centering
        \includegraphics[width=\linewidth]{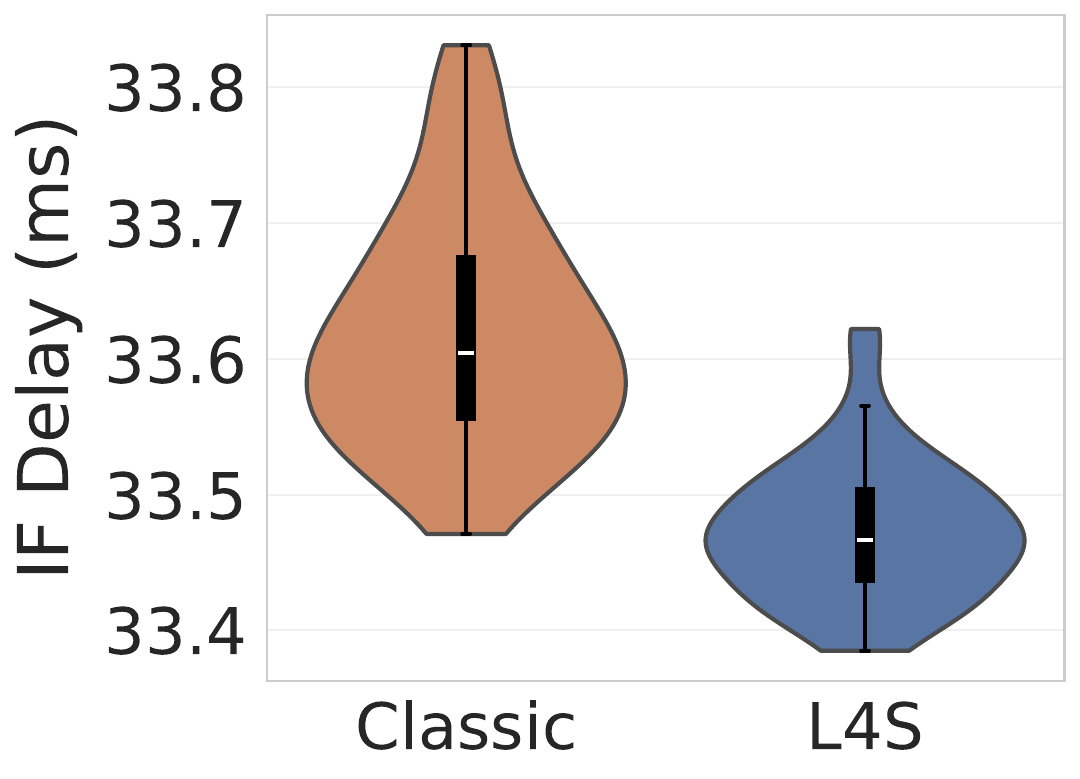}
        \captionsetup{font=footnotesize}
        \caption{Median IF Delay}
        \label{ifd-medmo_verizon_loss-0_rtt-20ms}
    \end{subfigure}
    \begin{subfigure}[b]{0.323\columnwidth}
        \centering
        \includegraphics[width=\linewidth]{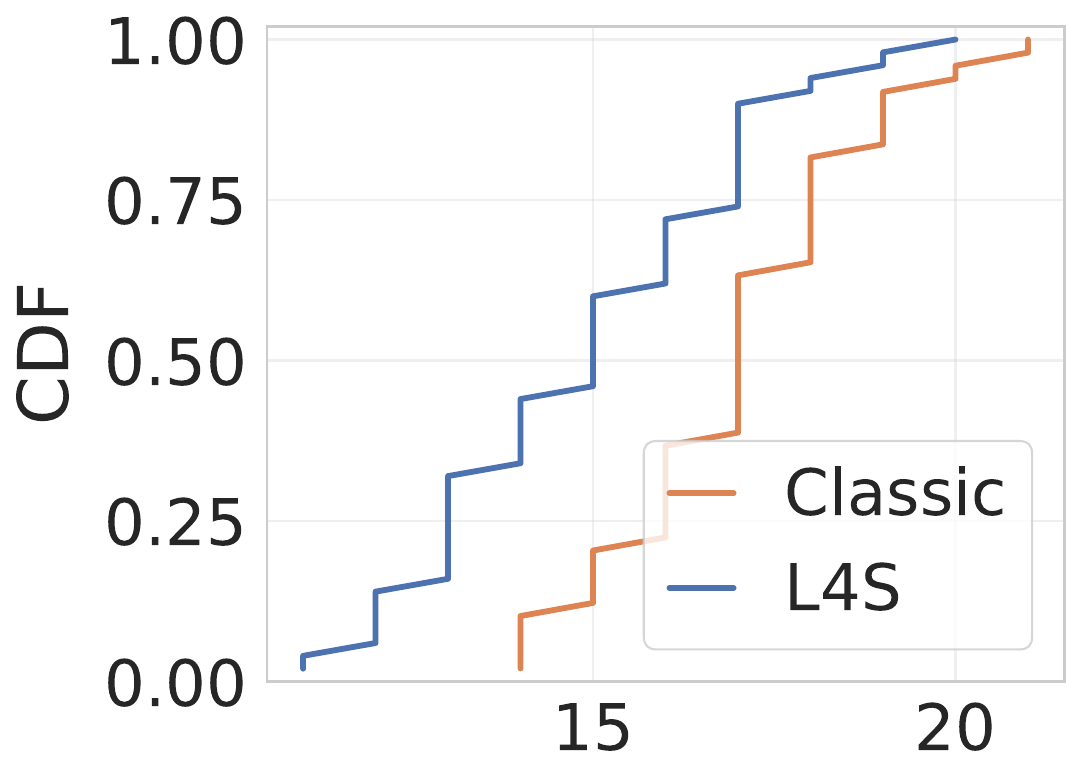}
        \captionsetup{font=footnotesize}
        \caption{Freeze Count (\#)}
        \label{fc-medmo_verizon_loss-0_rtt-20ms}
    \end{subfigure}
    \begin{subfigure}[b]{0.323\columnwidth}
        \centering
        \includegraphics[width=\linewidth]{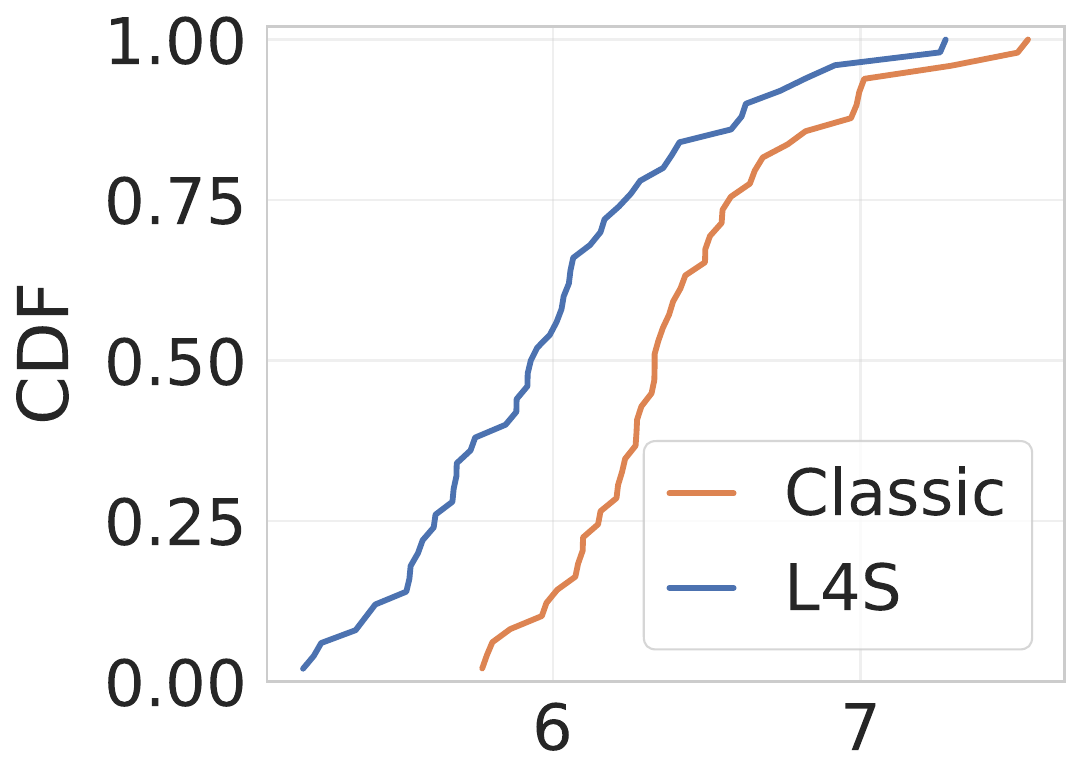}
        \captionsetup{font=footnotesize}
        \caption{Freeze Duration (s)}
        \label{fd-medmo_verizon_loss-0_rtt-20ms}
    \end{subfigure}

\caption{Classic vs. L4S for medium-motion video}
\label{fig:medmo_verizon_loss-0_rtt-20ms}
\end{figure}

To understand the effect of applying L4S to a video flow, we run the medium-motion video through the DualPI2 bottleneck over the Verizon LTE trace with 20 ms RTT. Fig.~\ref{fig:medmo_verizon_loss-0_rtt-20ms} shows the results for both Classic and L4S flows. \textbf{At the network level, L4S reduces the median per-run $95^{th}$ percentile queue delay from 35.8 ms in Classic to 23.1 ms, a 35.5\% reduction, but this comes with a 42.4\% drop in sender throughput, from about 3.0 Mbps to 1.7 Mbps.} In this baseline case, packet loss remains zero for both Classic and L4S, indicating that the observed differences are not due to drops. At the playback QoE level, the inter-frame (IF) delay changes only slightly: the median decreases from about 33.6 ms in Classic to roughly 33 ms in L4S, with L4S also exhibiting a slightly tighter distribution around the median. Freeze behavior improves more clearly under L4S. \textbf{The median freeze count decreases from 17 events in Classic to 15 in L4S, while the median freeze duration decreases from 6.33 s to 5.9 s, corresponding to reductions of about 12\% and 6\%, respectively.}

\begin{figure}[ht]
\centering
    \begin{subfigure}[b]{0.323\columnwidth}
        \centering
        \includegraphics[width=\linewidth]{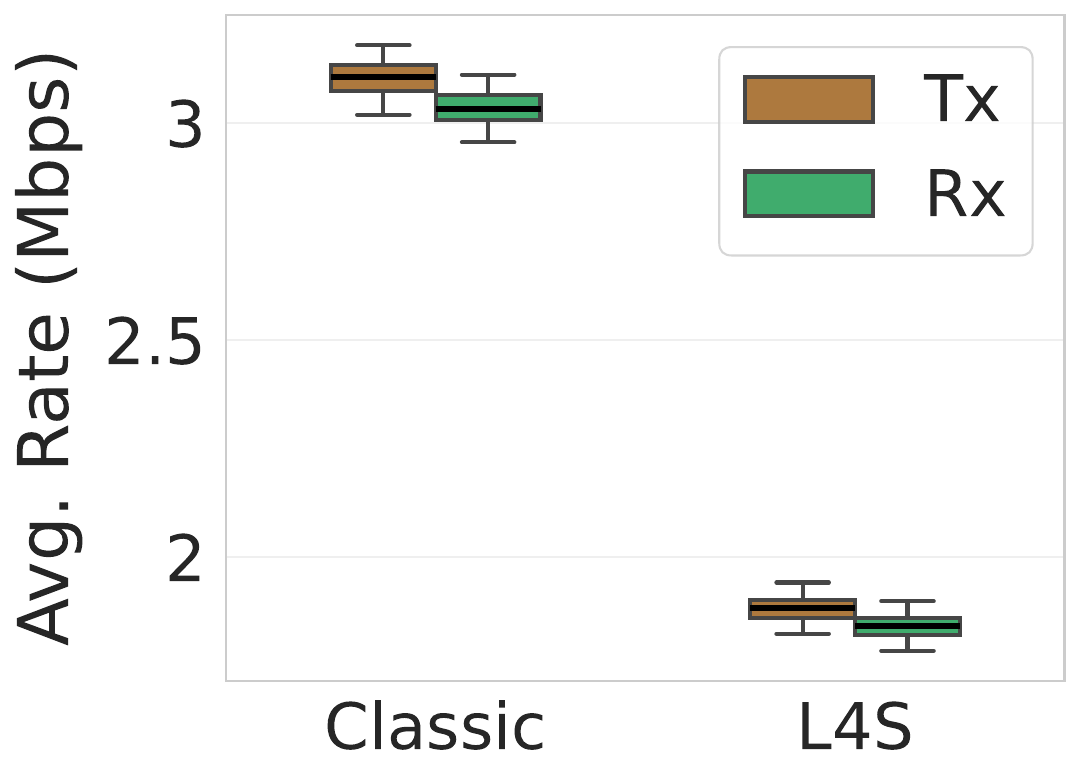}
        \captionsetup{font=footnotesize}
        \caption{Tx/Rx Rate}
    \end{subfigure}
    \begin{subfigure}[b]{0.323\columnwidth}
        \centering
        \includegraphics[width=\linewidth]{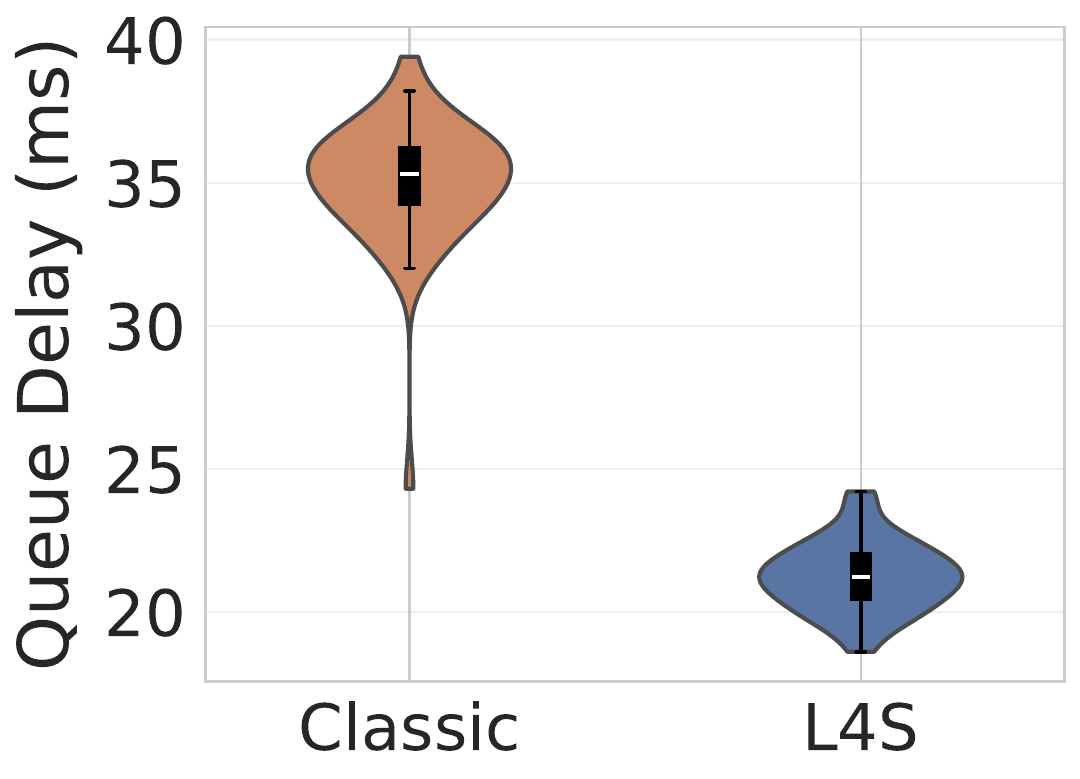}
        \captionsetup{font=footnotesize}
        \caption{Queue Delay (P95)}
    \end{subfigure}
    \begin{subfigure}[b]{0.323\columnwidth}
        \centering
        \includegraphics[width=\linewidth]{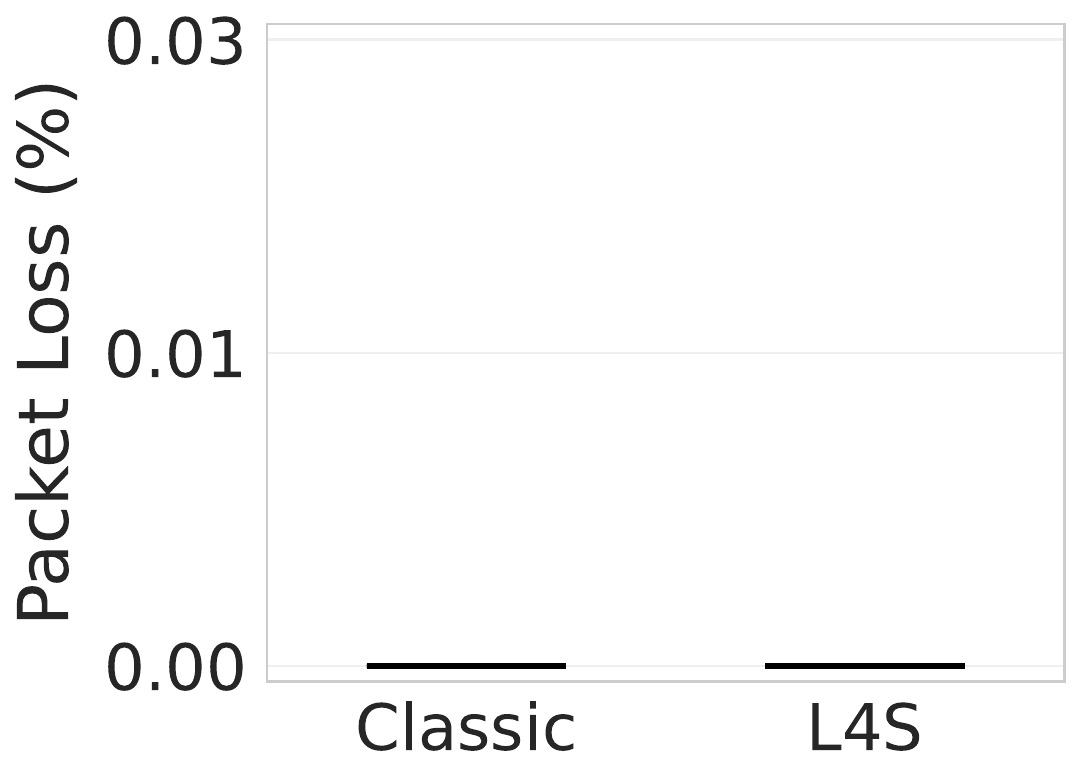}
        \captionsetup{font=footnotesize}
        \caption{Packet Loss}
    \end{subfigure}
\caption{Classic vs. L4S for synthetic RTP traffic}
\label{fig:fake_verizon_loss-0_rtt-20ms}
\end{figure}

\noindent \textbf{Sanity check against synthetic traffic.} Adding video encoding to the BW tool raises the question of whether the video pipeline preserves SCReAM’s original transport-level behavior. To answer this, we repeat the main scenario using the BW tool’s original fake RTP traffic mode and summarize the resulting network-level metrics in Fig.~\ref{fig:fake_verizon_loss-0_rtt-20ms}. The Tx/Rx rates, queue delay, and packet loss closely match those of the video-traffic case, showing that \textbf{the encoder-integration preserves, at the network metric level, the intended SCReAM control behavior seen with synthetic traffic for both Classic and L4S flows.} 

\subsection{Effect of random packet loss}
We introduce a 1\% random packet loss through Mahimahi. This setting introduces measurable loss and allows us to compare Classic and L4S under non-negligible drop conditions.

\begin{figure}[ht]
\centering
    \begin{subfigure}[b]{0.32\columnwidth}
        \centering
        \includegraphics[width=\linewidth]{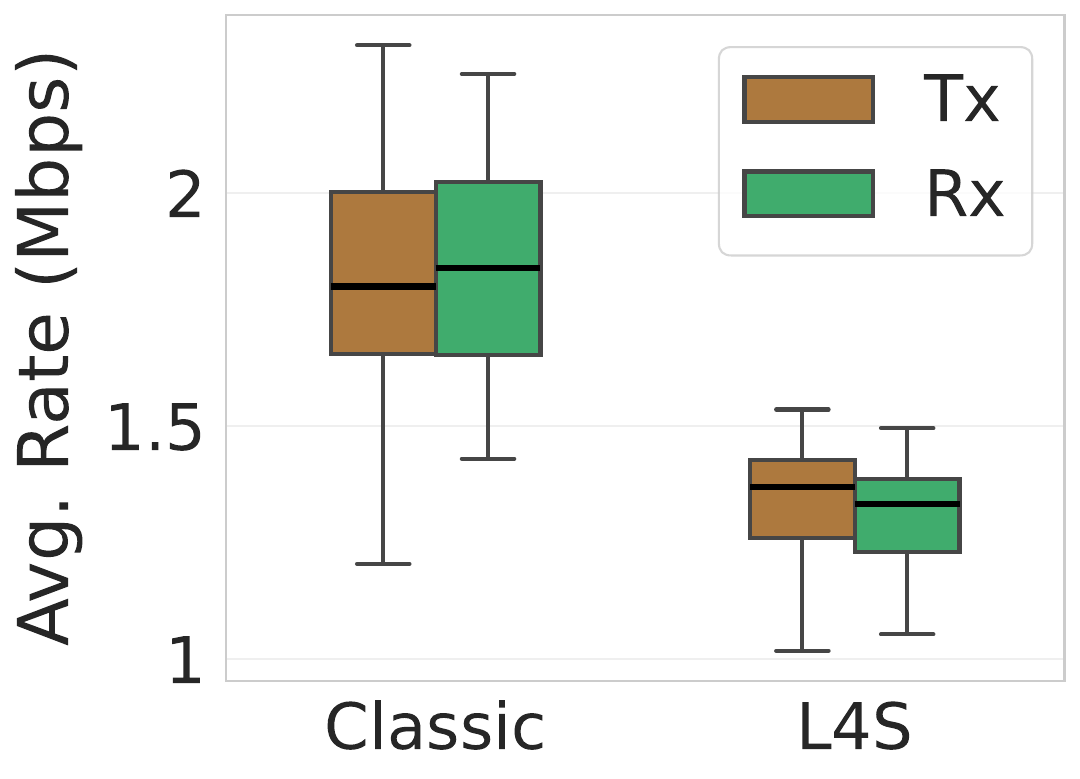}
        \captionsetup{font=footnotesize}
        \caption{Tx/Rx Rate}
    \end{subfigure}
    \hfill
    \begin{subfigure}[b]{0.32\columnwidth}
        \centering
        \includegraphics[width=\linewidth]{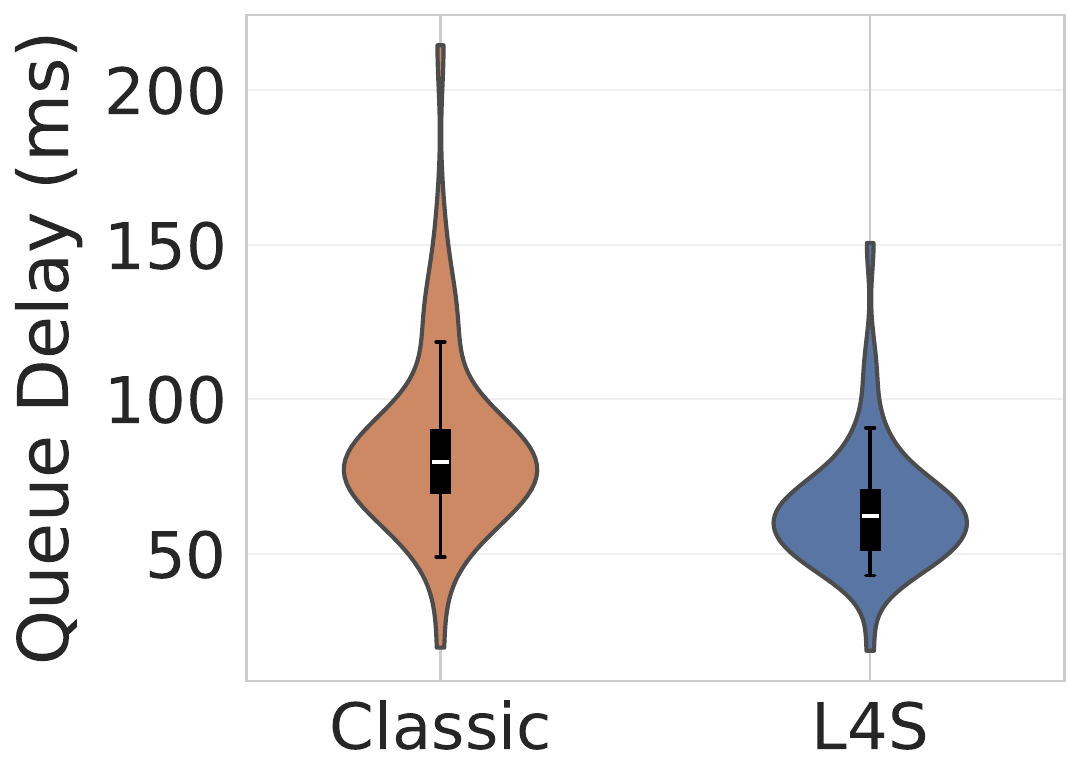}
        \captionsetup{font=footnotesize}
        \caption{Queue Delay (P95)}
    \end{subfigure}
    \hfill
    \begin{subfigure}[b]{0.32\columnwidth}
        \centering
        \includegraphics[width=\linewidth]{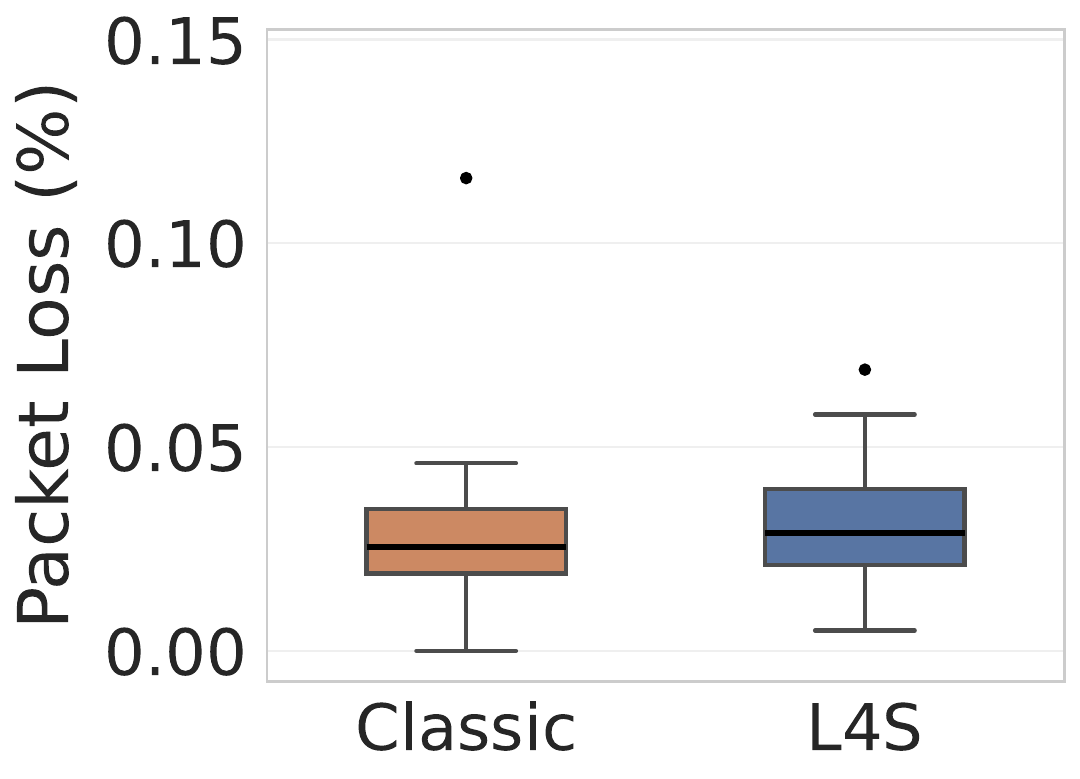}
        \captionsetup{font=footnotesize}
        \caption{Packet Loss}
    \end{subfigure}

    \vspace{0.5em}

    \begin{subfigure}[b]{0.32\columnwidth}
        \centering
        \includegraphics[width=\linewidth]{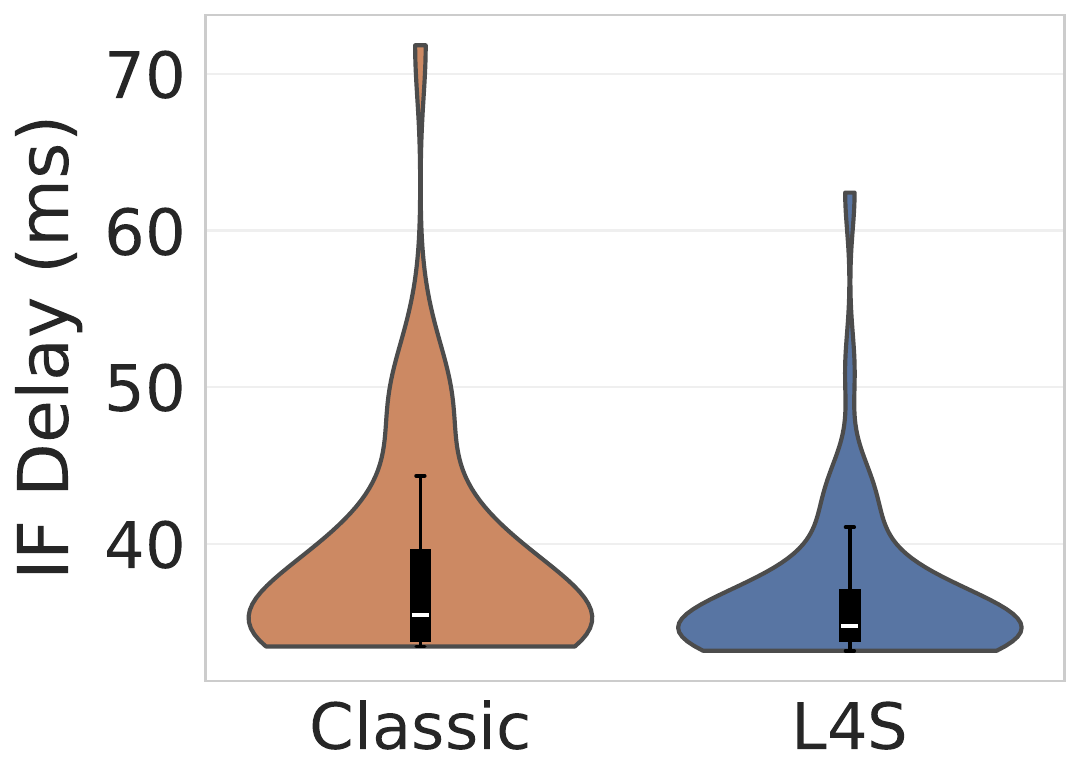}
        \captionsetup{font=footnotesize}
        \caption{Median IF Delay}
    \end{subfigure}
    \hfill
    \begin{subfigure}[b]{0.32\columnwidth}
        \centering
        \includegraphics[width=\linewidth]{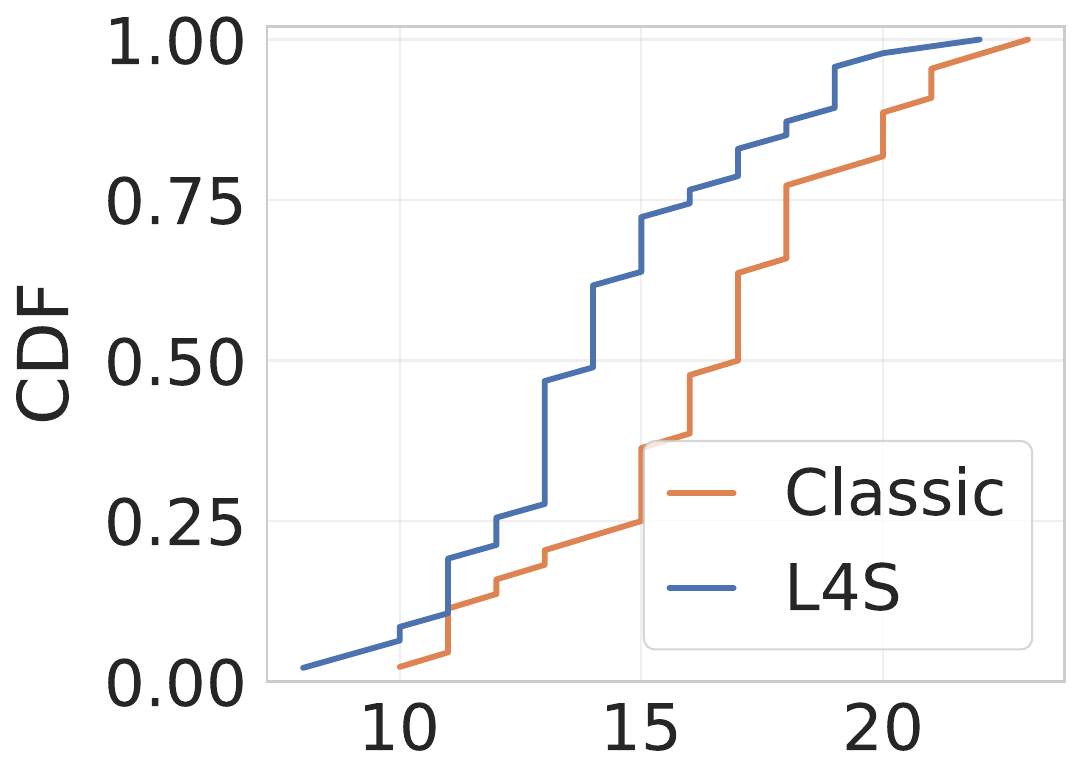}
        \captionsetup{font=footnotesize}
        \caption{Freeze Count (\#)}
    \end{subfigure}
    \hfill
    \begin{subfigure}[b]{0.32\columnwidth}
        \centering
        \includegraphics[width=\linewidth]{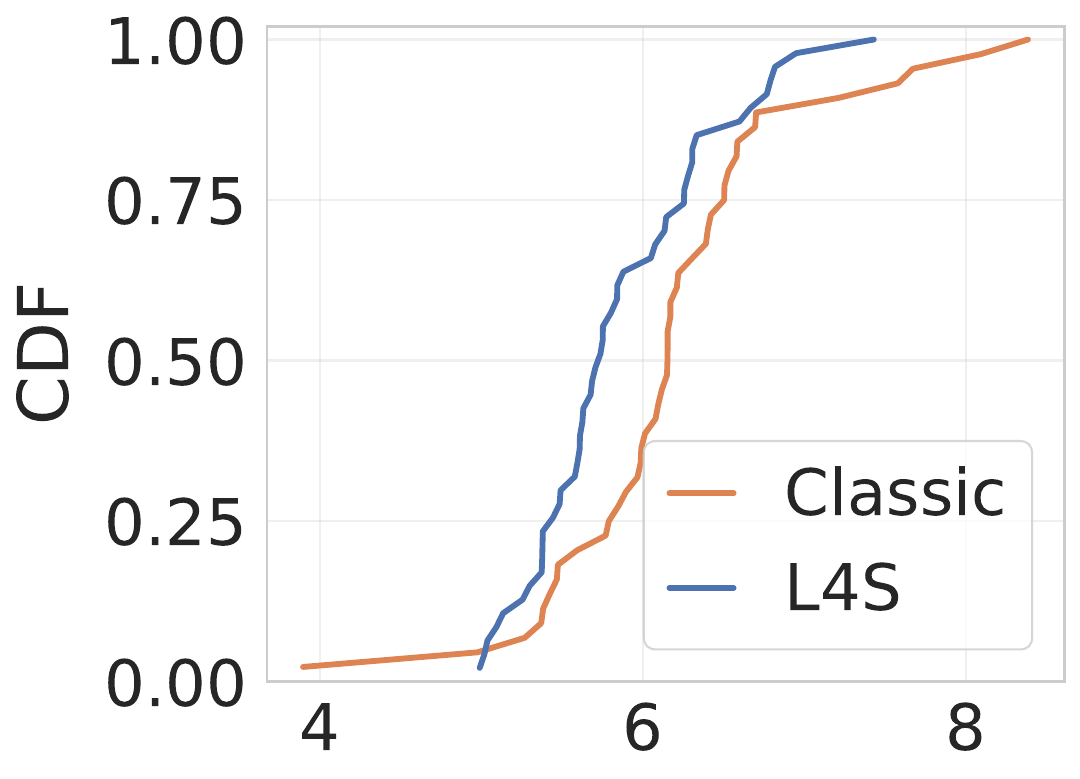}
        \captionsetup{font=footnotesize}
        \caption{Freeze Duration (s)}
    \end{subfigure}

\caption{Classic vs. L4S for medium-motion video with a 1\% packet loss rate}
\label{fig:medmo_verizon_loss-0.01_rtt-20ms}
\end{figure}

\noindent Fig.~\ref{fig:medmo_verizon_loss-0.01_rtt-20ms} shows that \textbf{L4S still reduces the median per-run $95^{th}$ percentile queue delay, from 79.7 ms in Classic to 62.1 ms, a 22\% reduction, while the sender throughput decreases from 1.8 Mbps to 1.3 Mbps, a 27.7\% drop.} Compared to the no-loss scenario, the gap between Classic and L4S narrows in terms of the queue delay as well as throughput, while the spread of these metrics increases, especially in the Classic mode.

At the sender, the detected packet loss remains negligible in both regimes and is slightly higher under L4S, with a median of 0.029\% versus 0.026\% for Classic. Both values remain far below the injected 1\% link loss, reflecting the effect of retransmissions.

At the playback level, the median IF delay is identical in both cases at 35 ms, but \textbf{L4S exhibits a narrower interquartile range (34--37 ms vs.\ 34--40 ms), indicating a more concentrated distribution and more stable playback across runs.} The median freeze count decreases from 17 in Classic to 14 in L4S, corresponding to a 17.6\% reduction. The median freeze duration decreases from 6.15 s in Classic to 5.7 s in L4S, a reduction of about 7\%. Both reductions are larger than in the baseline lossless scenario.

Overall, \textbf{while the network-level gap between Classic and L4S narrows in the presence of random loss, the QoE advantage of L4S becomes more pronounced.} This suggests a greater benefit of SCReAM in L4S mode under fluctuating, loss-prone mobile network conditions, and reinforces the need to analyze QoE alongside network metrics to capture the full application-level impact.

\subsection{Effect of video motion complexity}
\begin{figure}[ht]
\centering
    \begin{subfigure}[b]{0.48\columnwidth}
        \centering
        \includegraphics[width=\linewidth]{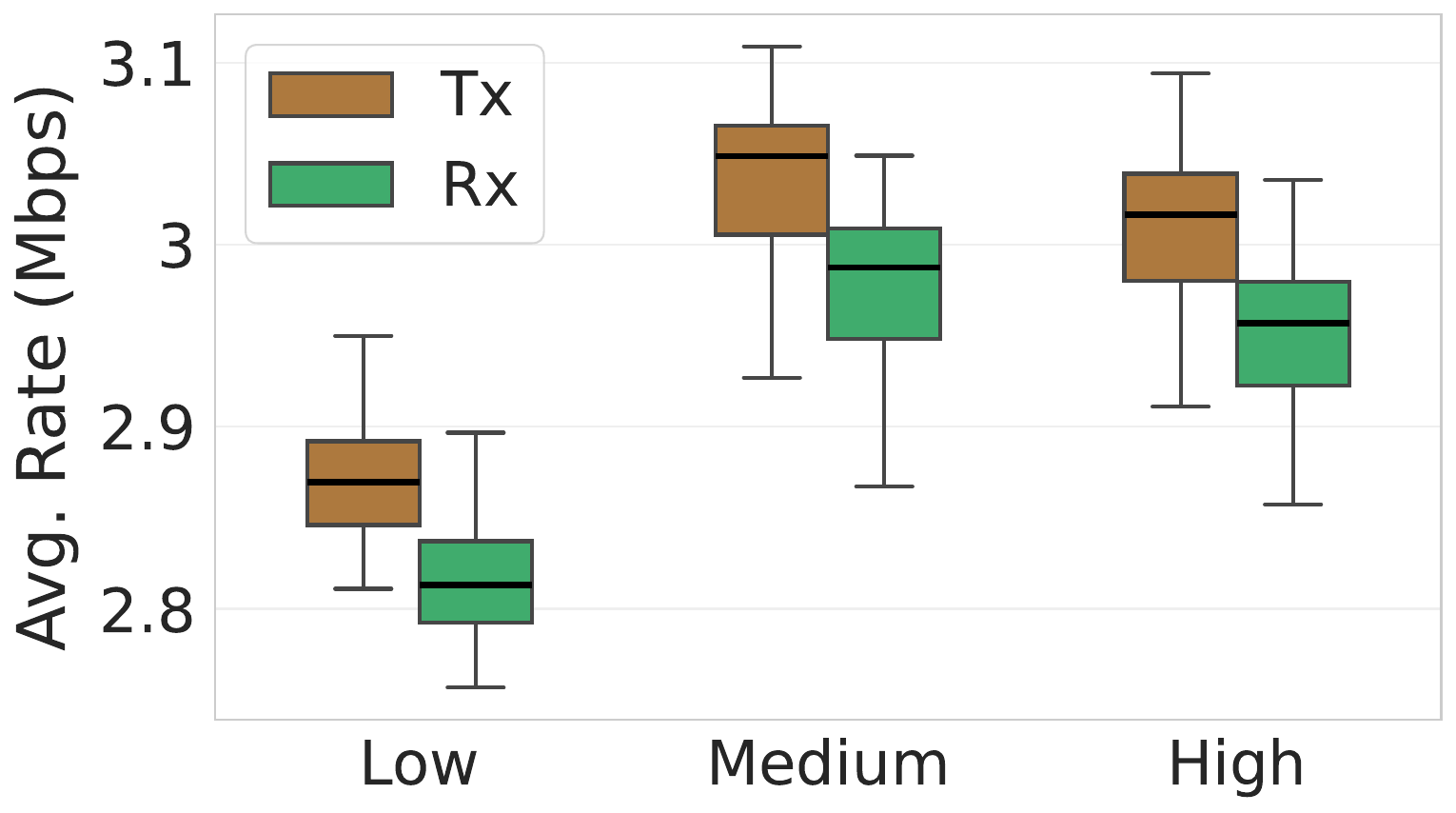}
        \captionsetup{font=footnotesize}
        \caption{Classic Tx/Rx Rate}
    \end{subfigure}
    \begin{subfigure}[b]{0.48\columnwidth}
        \centering
        \includegraphics[width=\linewidth]{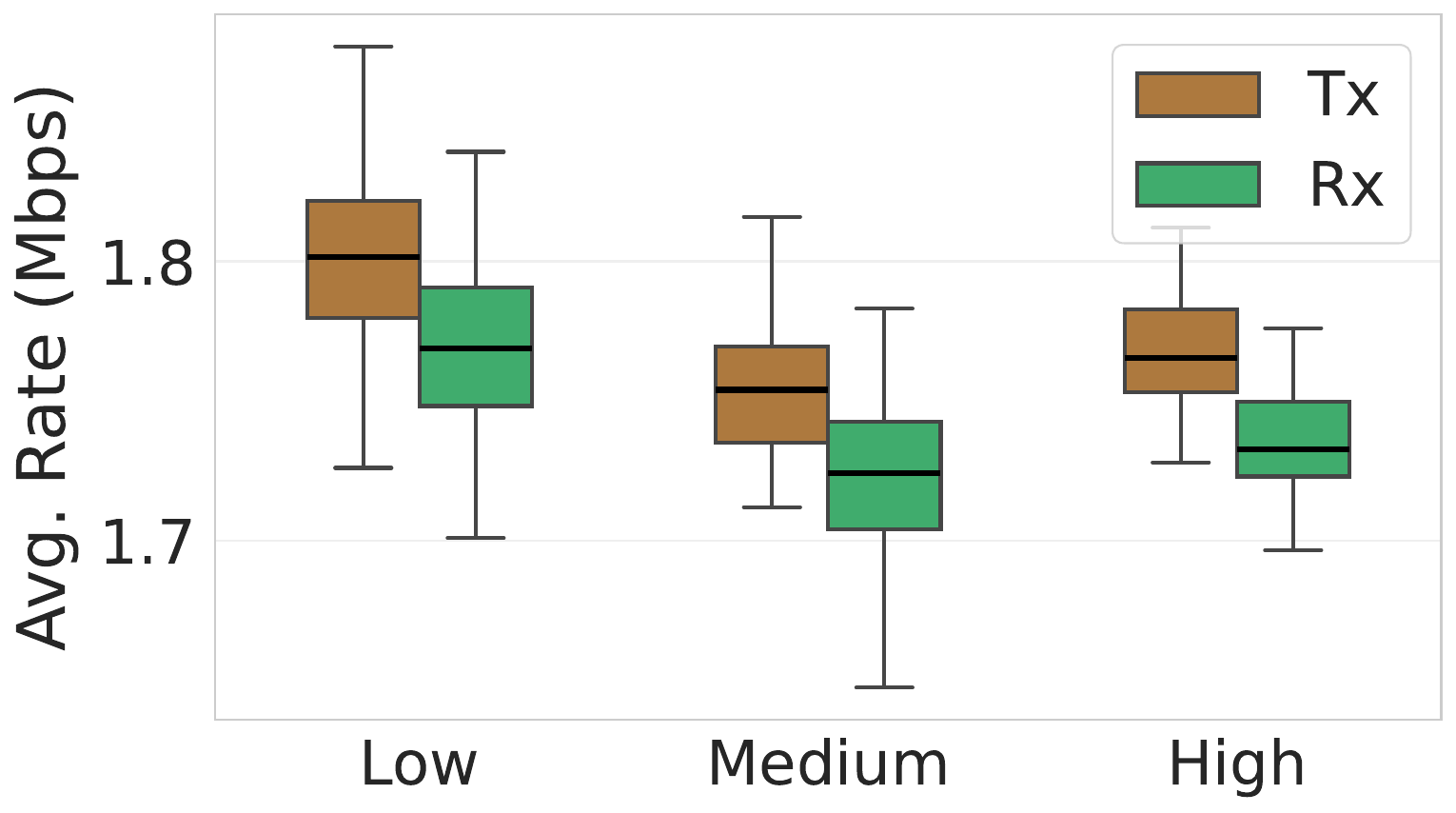}
        \captionsetup{font=footnotesize}
        \caption{L4S Tx/Rx Rate}
    \end{subfigure}
\caption{Tx/Rx rates for different motion levels}
\label{fig:all_tx_rx}
\end{figure}

In this experiment, we use the same Verizon LTE trace with a 20 ms RTT and no loss. Fig.~\ref{fig:all_tx_rx} summarizes the transmission and reception rates obtained for the different motion-complexity regimes. We notice that the Tx/Rx rate does not necessarily correlate monotonically with the motion complexity level. Indeed, higher-motion content does not automatically translate into a higher transmitted rate, since the encoder can react to increased temporal complexity by compressing more aggressively, restraining the output rate. As a result, content complexity is filtered through the joint behavior of the encoder and SCReAM, rather than being reflected directly in the network load.

\begin{figure}[ht]
\centering
    \begin{subfigure}[b]{0.48\columnwidth}
        \centering
        \includegraphics[width=\linewidth]{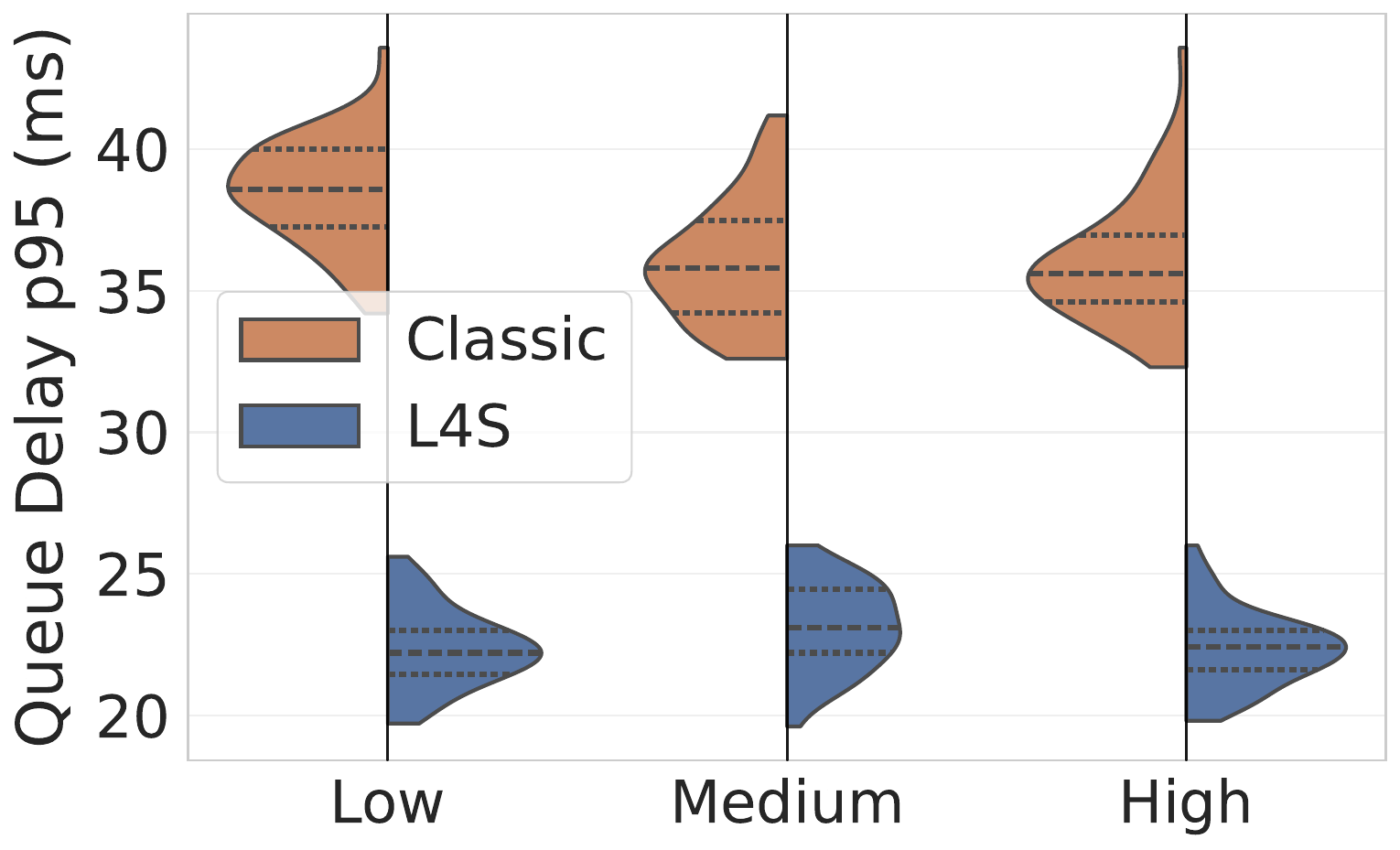}
        \captionsetup{font=footnotesize}
        \caption{Queue Delay (P95)}
        \label{fig:qd-all_motion}
    \end{subfigure}  
    \hfill
    \begin{subfigure}[b]{0.48\columnwidth}
        \centering
        \includegraphics[width=\linewidth]{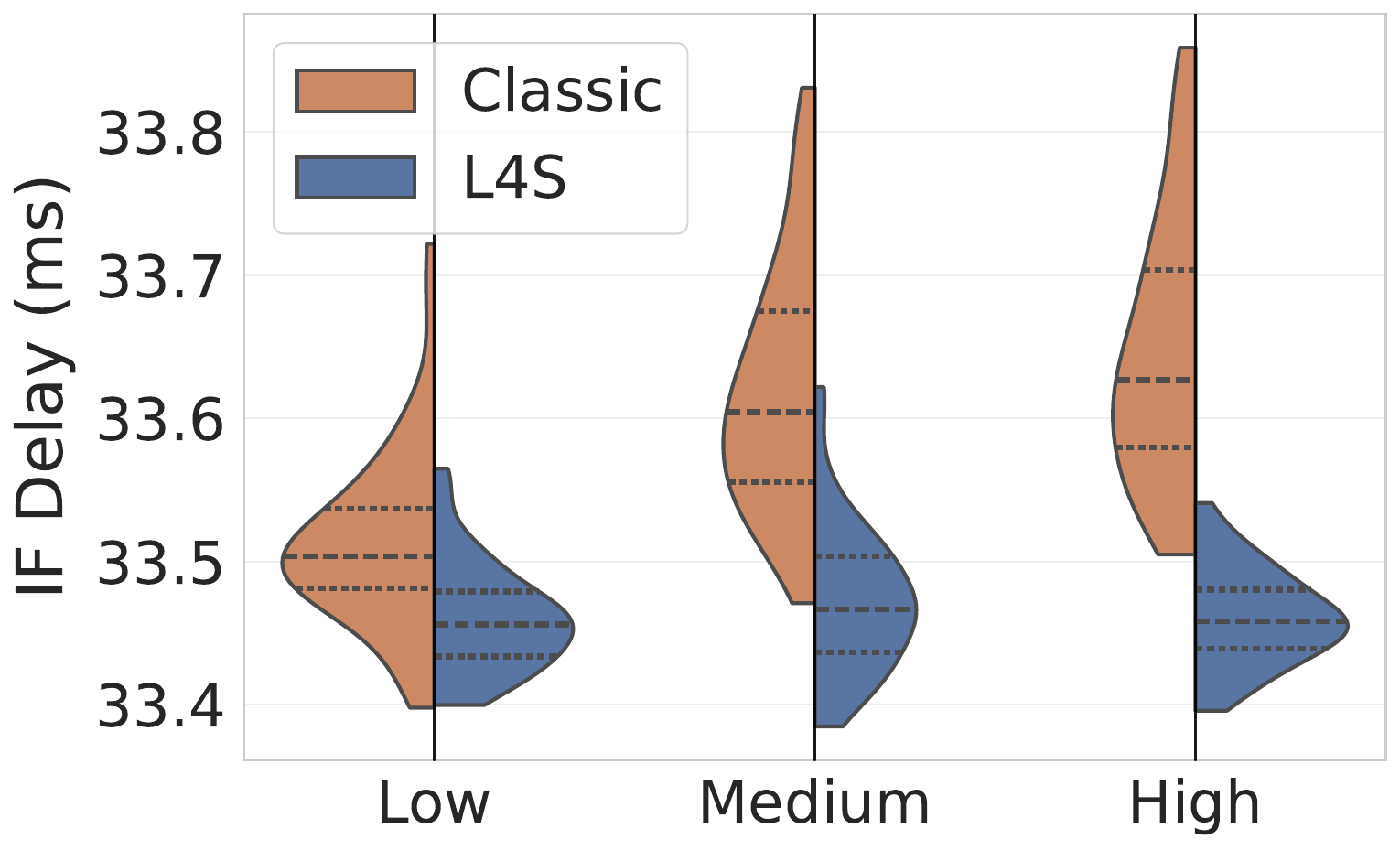}
        \captionsetup{font=footnotesize}
        \caption{Median IF Delay}
        \label{fig:ifd-all_motion_qoe}
    \end{subfigure}
\caption{P95 queuing delay and median IF delay for different motion levels}
\label{fig:all_motion_qoe}
\end{figure}

However, we can see from Fig.~\ref{fig:qd-all_motion} that \textbf{SCReAM in L4S mode shows higher stability in queuing delay than Classic across all motion levels, which translates into a more stable median IF delay,} as shown in Fig.~\ref{fig:ifd-all_motion_qoe}.

\section{Conclusion}

We extended the SCReAM BW tool with a video codec, enabling experiments with video traffic in addition to its original synthetic RTP traffic mode. Using this extension with Mahimahi, we evaluated SCReAM over Classic and L4S modes on a mobile network trace and under random packet loss, considering both network-level and QoE metrics.

We find that L4S reduces the median per-run $95^{th}$ percentile queue delay by 35\%, at the cost of a 42.4\% drop in sender throughput. Under a 1\% random loss, the network-level gap between Classic and L4S narrows, but the QoE advantage of L4S becomes more pronounced, with the reduction in freeze count increasing from 12\% in the lossless scenario to 17\% with 1\% random loss. Across different motion-complexity levels, L4S also shows greater stability than Classic in terms of median inter-frame delay. These results also highlight the importance of evaluating L4S not only through network metrics, but also through application-level QoE metrics on actual video traffic.

\begin{acks}
We thank Dr. Francis Yan for his sustained guidance and valuable feedback throughout the development of this work.   
\end{acks}

\printbibliography

\end{document}